\documentclass[a4paper,12pt]{article}
\usepackage[cp1251]{inputenc}
\usepackage[english]{babel}
\usepackage[dvips]{graphicx}
\graphicspath{{.}}
\usepackage{amsfonts}
\usepackage{amsmath}

\begin{document}
\mathsurround=2pt \sloppy
\title{\bf  Linear theory of random textures  of $^3$He-A in aerogel}
\author{I. A. Fomin
\vspace{.5cm}\\
{\it P. L. Kapitza Institute for Physical Problems}\\ {\it Russian
Academy of Science},\\{\it Kosygina 2,
 119334 Moscow, Russia}}
\maketitle
\begin{abstract}
Spacial variation of the orbital part of the order parameter of $^3$He-A in aerogel is represented as a random walk of the
unit vector $\mathbf{l}$ in a field of random anisotropy produced by the strands of aerogel.
For a range of distances, where variation of $\mathbf{l}$ is small in comparison with its absolute value correlation function of directions of $\mathbf{l}(\mathbf{r})$ is expressed in terms of the correlation function of the random anisotropy field. With simplifying assumptions about this correlation function a spatial dependence of the average variation $\langle\delta\mathbf{l}^2\rangle$ is found analytically for isotropic and axially anisotropic aerogels. Average projections of $\mathbf{l}$  on the axes of anisotropy are expressed in terms of characteristic parameters of the problem.
Within the ``model of random cylinders'' numerical estimations of characteristic length for disruption of the long-range order and of the critical anisotropy for restoration of this order are made and compared  with other estimations .

\end{abstract}

\section{}
 The order parameter of the A-phase of superfluid $^3$He has a form:
$$
A_{\mu j}=\Delta\frac{1}{\sqrt{2}}\hat d_{\mu}(\hat m_j+i\hat
n_j).                                                           \eqno(1)
$$
In the bulk liquid it is continuously degenerate with respect to separate rotations of its spin part - unit vector $\hat d_{\mu}$ and of its orbital part $\hat m_j+i\hat n_j$, where $\hat m_j$ and $\hat n_j$  are two mutually orthogonal
unit vectors. Usually these two vectors are appended by the third orbital vector $\mathbf{l}=\mathbf{m}\times\mathbf{n}$
to form an orthogonal triad. Orbital degeneracy is partly lifted by walls of a container or by extraneous objects.
Aerogel when immersed in $^3$He-A creates a random local anisotropy, which orients the orbital part of the order parameter. The anisotropy  varies on a length scale of the order of the average distance $\xi_c$ between the strands of aerogel. Following the argument of Larkin \cite{Lark} and Imry and Ma \cite{Ima} Volovik \cite{vol1} predicted that the random anisotropy disrupts the long-range order in $^3$He-A and creates a glass-like Larkin-Imry-Ma (LIM) state. It is a random texture, in which the form of the order parameter, given by Eq. (1) locally does not change, but orientation of its orbital part varies on a length scale $\xi_{LIM}$, which is determined by an interplay of spatial fluctuations of the random anisotropy and the gradient energy of the condensate. If except for the local a regular global anisotropy is present, different random textures can realize.  They include the orbital glass (OG) state, the orbital ferromagnet (OF), the planar LIM state, and the spin-glass (SG) state where orientation of spin vector $\hat d_{\mu}$ is also random \cite{dm2}. The qualitative macroscopic argument \cite{vol2} renders an order of magnitude estimations for $\xi_{LIM}$ and for a minimal global anisotropy sufficient for a restoration of the long range order. Still  for a more detailed investigation of properties of random textures and for a quantitative interpretation of experiments a more elaborate theoretical treatment may be required.

In a present paper only orbital textures are discussed.
A  proposed theory is based on calculation of correlation function for orientations of $\mathbf{l}(\mathbf{r})$ at two   points $\mathbf{r}_1$ and $\mathbf{r}_2$, which are so close that variation of  $\mathbf{l}(\mathbf{r})$ on a distance $|\mathbf{r}_2-\mathbf{r}_1|$ is small in comparison with its absolute value. Such approach is conceptually similar to that used in the theory of ferromagnets with a weak random anisotropy \cite{sasl}, but the results of this paper can not be transferred directly to the random textures of  $^3$He-A.   A principal reason is the essential difference of the order parameters of two systems, which in its turn effects a choice of problems and an approach to their solution.

Many observable properties of of  $^3$He-A (e.g. the NMR-shift, the anisotropy of the superfluid density) depend on orientation of the $\mathbf{l}$. The proposed theory makes possible to find a dependence of
$\langle\delta\mathbf{l}^2\rangle=\langle(\mathbf{l}(\mathbf{r}_2)-\mathbf{l}(\mathbf{r}_1))^2\rangle$ on a distance $|\mathbf{r}_2-\mathbf{r}_1|$ in a region where $\langle\delta\mathbf{l}^2\rangle\ll\mathbf{l}_0^2$. The angular brackets here and in what follows denote average over ensemble. Extrapolation of the obtained expression to $\langle\delta\mathbf{l}^2\rangle/\mathbf{l}_0^2\sim 1$ renders an alternative estimation for $\xi_{LIM}$ and for the critical global anisotropy, stabilizing the OF state. Both estimations  by the order of magnitude do not contradict to these made in Ref.\cite{vol2} with the aid of macroscopic argument. For anisotropic aerogels the theory renders also the average values of squares of projections of $\mathbf{l}$ on the axes of anisotropy. These averages determine the values of NMR frequency shifts in different states.

The LIM state of $^3$He-A in aerogel is also a good example of a \emph{pseudo-gap} state. The averaged density of states here is practically gapped but there is no superfluidity. Situation is reminiscent of that in high-T$_c$ superconductors. That increases interest in investigation of glass-like states of $^3$He-A.

\section{}
Formally interaction of aerogel with the order parameter of superfluid $^3$He is described by the additional term in the free energy density. This term is proportional to $\eta_{jl}({\bf r})A_{\mu j}A^*_{\mu l}$, where $\eta_{jl}({\bf r})$ is a random real symmetric traceless tensor. The isotropic part of interaction is included in the local suppression of the $T_c$, it affects the absolute value of the order parameter, but not its orientation.  Not too far from the transition temperature the equilibrium texture can be found by minimization of the Ginzburg and Landau functional. For high porosity aerogels  $\eta_{jl}({\bf r})$ can be treated as a perturbation. Only orientation dependent contributions to this functional are significant:
$$
 F_{GL}=N(0)\Delta^2\int d^3r\left\{\eta_{jl}({\bf r})\Delta_j\Delta^*_l+\xi_s^2\left(|rot\vec{\Delta}|^2+
3|div\vec{\Delta}|^2\right)\right\},                  \eqno(2)
$$
where $\vec{\Delta}=\frac{1}{\sqrt{2}}(\mathbf{m}+i\mathbf{n})$, $\xi_s^2=\frac{7\zeta(3)}{20}\xi_0^2$,  $\xi_0^2=\frac{\hbar v_F}{2\pi T_c}$  \cite{VW}.
Variation of the functional (2) with respect to orientation of the triad $(\mathbf{m},\mathbf{n},\mathbf{l})$, according to
$\delta\mathbf{m}=\vec{\theta}\times\mathbf{m}$ etc., where $\vec{\theta}$ is an infinitesimal rotation vector, renders the equation determining the equilibrium texture:
$$
\mathbf{l}\times\overrightarrow{\eta\mathbf{l}}+\xi_s^2[\mathbf{m}\times(2\nabla (\nabla\cdot\mathbf{m})+\nabla^2\mathbf{m})+
\mathbf{n}\times(2\nabla(\nabla\cdot\mathbf{n})+\nabla^2\mathbf{n})]=0.             \eqno(3)
$$
Taking projections of this equation on each of the directions $\mathbf{m},\mathbf{n},\mathbf{l}$ we arrive at three scalar equations:
$$
\mathbf{n}\cdot(\overrightarrow{D\mathbf{m}})=\mathbf{m}\cdot(\overrightarrow{D\mathbf{n}}),         \eqno(4)
$$
$$
\mathbf{l}\cdot(\overrightarrow{D\mathbf{m}})=\mathbf{m}\cdot(\overrightarrow{\eta\mathbf{l}}),       \eqno(5)
$$
$$
\mathbf{l}\cdot(\overrightarrow{D\mathbf{n}})=\mathbf{n}\cdot(\overrightarrow{\eta\mathbf{l}}),       \eqno(6)
$$
where shorthand notations $\overrightarrow{D\mathbf{m}}=\xi_s^2[2\nabla (\nabla\cdot\mathbf{m})+\nabla^2\mathbf{m}]$ and
$\overrightarrow{D\mathbf{n}}=\xi_s^2[2\nabla (\nabla\cdot\mathbf{n})+\nabla^2\mathbf{n}]$ are used.
In a zero order approximation over $\eta_{jl}$ arbitrarily oriented spatially homogeneous triad $(\mathbf{m},\mathbf{n},\mathbf{l})$ satisfies Eqns. (4)-(6). This solution is continuously degenerate. Consider first an isotropic aerogel $\langle\eta_{jl}({\bf r})\rangle=0$. At the strength of the above mentioned  LIM argument  even arbitrary small random perturbation disrupts the orientational long-range order, preserving orientation of the triad only over a finite distance of the order of the LIM correlation length $\xi_{LIM}$ which grows when $\eta_{jl}$  tends to zero.  Random anisotropy $\eta_{jl}$ varying over a distance of the order of $\xi_c$ induces a random walk of vector $\mathbf{l}$ over the sphere $\mathbf{l}^2=1$.  Since $\eta_{jl}$ is a perturbation the resulting  $\xi_{LIM}\gg \xi_0$. For two points $\mathbf{r}_1$ and $\mathbf{r}_2$ meeting the condition $|\mathbf{r}_2-\mathbf{r}_1|\ll\xi_{LIM}$ average variation  $\langle(\mathbf{l}(\mathbf{r}_2)-\mathbf{l}(\mathbf{r}_1))^2\rangle\ll \mathbf{l}_0^2$. Within a region $\xi_c\ll |\mathbf{r}_2-\mathbf{r}_1|\ll\xi_{LIM}$ one can introduce an average orientation of the triad $\mathbf{m}_0,\mathbf{n}_0,\mathbf{l}_0$, so that e.g. $\mathbf{l}(\mathbf{r}_1)=\mathbf{l}_0+\vec{\lambda}(\mathbf{r}_1)$  and $\mathbf{l}(\mathbf{r}_2)=\mathbf{l}_0+\vec{\lambda}(\mathbf{r}_2)$ etc.. Both $\vec{\lambda}(\mathbf{r}_1)$ and $\vec{\lambda}(\mathbf{r}_2)$ are small. With account of the normalization condition for $\mathbf{l}(\mathbf{r})$ the increment $\vec{\lambda}(\mathbf{r})$ can be represented as
  $\vec{\lambda}(\mathbf{r})=\vec{\theta}(\mathbf{r})\times\mathbf{l}_0$, where $\vec{\theta}(\mathbf{r})$ is a vector of small rotation. The same $\vec{\theta}(\mathbf{r})$ determines rotation of all triad  $(\mathbf{m},\mathbf{n},\mathbf{l})$: $\mathbf{m}(\mathbf{r})=\mathbf{m}_0+\vec{\mu}(\mathbf{r})$, $\vec{\mu}(\mathbf{r})=\vec{\theta}(\mathbf{r})\times\mathbf{m}_0$ and $\mathbf{n}(\mathbf{r})=\mathbf{n}_0+\vec{\nu}(\mathbf{r})$, $\vec{\nu}(\mathbf{r})=\vec{\theta}(\mathbf{r})\times\mathbf{n}_0$.

Of a principal interest is the average $\langle(\mathbf{l}(\mathbf{r}_2)-\mathbf{l}(\mathbf{r}_1))^2\rangle=
\langle(\vec{\lambda}(\mathbf{r}_2)-\vec{\lambda}(\mathbf{r}_1))^2\rangle=
2\langle(\vec{\lambda}(\mathbf{r}_1))^2-\vec{\lambda}(\mathbf{r}_1)\cdot\vec{\lambda}(\mathbf{r}_2)\rangle$,
it can be expressed in terms of $\vec{\theta}(\mathbf{r})$:  $\langle(\mathbf{l}(\mathbf{r}_2)-\mathbf{l}(\mathbf{r}_1))^2\rangle=
2\langle\vec{\theta}_{\perp}(\mathbf{r}_1)\vec{\theta}_{\perp}(\mathbf{r}_1)-
\vec{\theta}_{\perp}(\mathbf{r}_2)\vec{\theta}_{\perp}(\mathbf{r}_1)\rangle \mathbf{l}_0^2$,
 where $\vec{\theta}_{\perp}(\mathbf{r}_2)\vec{\theta}_{\perp}(\mathbf{r}_1)=\vec{\theta}(\mathbf{r}_2)\cdot\vec{\theta}(\mathbf{r}_1)-
(\vec{\theta}(\mathbf{r}_2)\cdot\mathbf{l}_0)(\vec{\theta}(\mathbf{r}_1)\cdot\mathbf{l}_0)/\mathbf{l}_0^2$. $\vec{\theta}_{\perp}(\mathbf{r})$ is the projection of  $\vec{\theta}(\mathbf{r})$ on a plane, normal to $\mathbf{l}_0$. To find a first order solution of Eqns. (4)-(6) we have to substitute the increments $\vec{\mu}(\mathbf{r}),\vec{\nu}(\mathbf{r}),\vec{\lambda}(\mathbf{r})$ expressed in terms of $\vec{\theta}(\mathbf{r})$ in Eqns. (4)-(6) and linearize these equations over $\vec{\theta}(\mathbf{r})$.
In the coordinate system with the $x,y,z$ axes directed respectively along $\mathbf{m}_0,\mathbf{n}_0,\mathbf{l}_0$ the linearized equations are:
$$
\nabla^2\theta_x+2\frac{\partial}{\partial z}\left(\frac{\partial \theta_x}{\partial z}-\frac{\partial\theta_z}{\partial x}\right) =\frac{\eta_{yz}}{\xi_s^2}.              \eqno(7)
$$
$$
\nabla^2\theta_y+2\frac{\partial}{\partial z}\left(\frac{\partial \theta_y}{\partial z}-\frac{\partial \theta_z}{\partial y}\right) =-\frac{\eta_{xz}}{\xi_s^2},              \eqno(8)
$$
$$
2\nabla^2\theta_z-\frac{\partial}{\partial z}div\vec{\theta}=0,              \eqno(9)
$$
These equations  can be solved by Fourier transformation:$\theta_{x,y,z}(\mathbf{r})=\int\exp(i\mathbf{k}\mathbf{r})\theta_{x,y,z}(\mathbf{k})\frac{Vd^3k}{(2\pi)^3}$, where $V$ is a normalization volume:
$$
\theta_x=\frac{1}{k^2+2k_z^2}[\hat{k}_x\hat{k}_y\eta_{xz}-(1+\hat{k}^2_x)\eta_{yz}],        \eqno(10)
$$
$$
\theta_y=\frac{1}{k^2+2k_z^2}[(1+\hat{k}^2_y)\eta_{xz}-\hat{k}_x\hat{k}_y\eta_{yz}],        \eqno(11)
$$
$$
\theta_z=\frac{1}{2k^2+k_z^2}[\hat{k}_z\hat{k}_y\eta_{xz}-\hat{k}_z\hat{k}_x\eta_{yz}].        \eqno(12)
$$

Only $\theta_x$ and $\theta_y$ enter expression for  $\langle(\mathbf{l}(\mathbf{r}_2)-\mathbf{l}(\mathbf{r}_1))^2\rangle\equiv
\langle[\delta\mathbf{l}(\mathbf{r})]^2\rangle$, where $\mathbf{r}=\mathbf{r}_2-\mathbf{r}_1$.
Interaction term in the functional (2) can be rewritten as $-\eta_{jl}l_jl_l$. It contains only  $\mathbf{l}$.
 Then,  in a principal order over $\eta_{jl}$
$$
\frac{\langle[\delta\mathbf{l}(\mathbf{r})]^2\rangle}{\mathbf{l}_0^2}=\int[1-\exp(i\mathbf{k}\cdot\mathbf{r})]
\frac{K(\mathbf{k})}{\xi_s^4(k_x^2+k_y^2+
\gamma^2k_z^2)^2}\frac{d^3k}{(2\pi)^3}.                                 \eqno(13)
$$
The numerator  $K(\mathbf{k})$ depends on correlation functions $\langle\eta_{xz}(-\mathbf{k})\eta_{xz}\rangle, \langle\eta_{yz}(-\mathbf{k})\eta_{xz}\rangle$ etc.. It can be represented as a sum of axially symmetric  $K_{s}(\mathbf{k})\equiv 2V\langle\eta_{xz}(-\mathbf{k})\eta_{xz}(\mathbf{k})+\eta_{yz}(-\mathbf{k})\eta_{yz}(\mathbf{k})\rangle$
and axially non-symmetric $K_{n}(\mathbf{k})\equiv 2V\langle(3-\hat{k}_z^2)[\hat{k}_y\eta_{xz}(\mathbf{k})-\hat{k}_x\eta_{yz}(\mathbf{k})]
[\hat{k}_y\eta_{xz}(-\mathbf{k})-\hat{k}_x\eta_{yz}(-\mathbf{k})]\rangle$ parts
($\gamma^2$ is the ratio of the ``longitudinal'' and ``transverse'' elasticities,  in our case it equals to 3).
When $\mathbf{l}(\mathbf{r})$ experiences a random walk $\langle[\delta\mathbf{l}(\mathbf{r})]^2\rangle$  grows with $\mathbf{r}$. Approximation Eq. (13) does apply until $\langle[\delta\mathbf{l}(\mathbf{r})]^2\rangle\ll \mathbf{l}_0^2$. Values of $\mathbf{r}$ corresponding to a boundary  of the region $\langle[\delta\mathbf{l}(\mathbf{r})]^2\rangle\sim \mathbf{l}_0^2$ can be used as an estimation of $\xi_{LIM}$. The rate of grows of $\langle[\delta\mathbf{l}(\mathbf{r})]^2\rangle$  depends on the explicit form of the function $K(\mathbf{k})$.
As an example of application of general formula we consider a simplified model expression for  $K(\mathbf{k})$. First, we assume that $K_{s}(\mathbf{k})\equiv K$ does not depend on $\mathbf{k}$. Second, we neglect contribution of $K_{n}(\mathbf{k})$.
Account of the axial non-symmetry does not change significantly  evaluation of $\langle[\delta\mathbf{l}(\mathbf{r})]^2\rangle$. It may become significant at discussion of superfluid properties of glass-like state.  For the simplified model
the integral in Eq. (13) can be done explicitly. The integration becomes trivial after a re-scaling of $z$-coordinate i.e. by the substitution instead of $\mathbf{r}$ of a new vector $\vec{\rho}=(x,y,z/\gamma)$ and respectively  $k_x=q_x,k_y=q_y,k_z=q_z/\gamma$. Using the value of the integral $\int_0^{\infty}\left(1-\frac{\sin y}{y}\right)\frac{dy}{y^2}=\frac{\pi}{4}$ we arrive at
$$
\frac{\langle[\delta\mathbf{l}(\mathbf{r})]^2\rangle}{\mathbf{l}_0^2}=\frac{K}{8\pi\gamma\xi_s^4}\rho=\frac{\rho}{\xi_{LIM}},                      \eqno(14)
$$
where $\xi_{LIM}=\frac{8\pi\gamma\xi_s^4}{K}$. Since $\rho$ is a distance, compressed along the direction of $\mathbf{l}_0$ the real disruption length is anisotropic -- the condition $\frac{\langle[\delta\mathbf{l}(\mathbf{r})]^2\rangle}{\mathbf{l}_0^2}\sim 1$ is met for $r\sim\xi_{LIM}$ if $\mathbf{r}$ is  perpendicular to $\mathbf{l}_0$ and for $r\sim\gamma\xi_{LIM}$ if they are parallel.

To be more specific we evaluate here $K(\mathbf{k})$ for an ensemble of randomly distributed and randomly oriented identical uniaxial impurities.  Following the theory of ``small objects'' in superfluid $^3$He \cite{Rainer} we find:
$$
\eta_{jl}(\mathbf{r})=\frac{\pi^2}{4}\xi_0\sum_s\left[\sigma^{(i)}\delta_{jl}+\sigma^{(a)}(3\hat{a}^s_j\hat{a}^s_l-\delta_{jl})\right]
\delta(\mathbf{r}-\mathbf{r}_s),          \eqno(15)
$$
where $\mathbf{r}_s$ are positions of impurities, $\hat{a}^{(s)}_j$-a unit vector in the direction of symmetry axis of the impurity at a point $\mathbf{r}_s$  and $\sigma_{tr}^{(i)}$ and $\sigma_{tr}^{(a)}$ are respectively isotropic and anisotropic parts of the transport cross-section for scattering of quasi-particles by the impurity defined in Refs.\cite{Rainer}. Substitution of Eq. (15) in the definition of $K(\mathbf{k})$ renders
$$
K(\mathbf{k})=\frac{1}{2V}\left(\frac{3\pi^2\xi_0}{2}\right)^2\langle\sum_{s,t}(\sigma_{tr}^{(a)})^2\hat{a}^s_x\hat{a}^s_z\hat{a}^t_x\hat{a}^t_z
\exp[i\mathbf{k}\cdot(\mathbf{r}_t-\mathbf{r}_s)]\rangle                 \eqno(16)
$$
If positions and orientations of different impurities are not correlated then in the sum over $s$ and $t$ only contribution of terms with $s=t$ remains finite after averaging. The angular averaging renders a factor $1/15$.  The resulting
$K(\mathbf{k})=\frac{3n}{40}(\pi^2\xi_0\sigma_{tr}^{(a)})^2$ does not depend on $\mathbf{k}$, it can be substituted in Eq. (14)
$$
\frac{\langle[\delta\mathbf{l}(\mathbf{r})]^2\rangle}{\mathbf{l}_0^2}=\frac{3\pi^2}{320}\frac{(\xi_0\sigma_{tr}^{(a)})^2}{\gamma\xi_s^4}
n\rho.                     \eqno(17)
$$
This result renders an estimation of the characteristic disruption length $\xi_{LIM}\sim\frac{320}{3\pi^3}\frac{\gamma\xi_s^4}{n(\xi_0\sigma_{tr}^{(a)})^2}$.
For a numerical estimation we assume that impurities are circular cylinders with the hight $b$ and radius $\varrho$. Then the  concentration $n$ can be expressed in terms of porosity $P$ via $n\pi\varrho^2 b=1-P$. Cross-section $\sigma_{tr}^{(a)}=-(\pi b\varrho)/8$ \cite{FS2}. It has to be remarked that characterization of contribution of an impurity to $\eta_{jl}(\mathbf{r})$ by its cross-section is justified when $b\ll\xi_0$, but as has been shown in Ref. \cite{SF1} for long cylinders only a part of their length of the order of $\xi_0$ renders significant contribution to the random anisotropy. For a practical estimation we introduce as an ``efficient'' hight $b=2\xi_0$. With these assumptions we arrive at $\xi_{LIM}\approx 10\frac{\xi_0}{1-P}$ For 98\% silica aerogel it renders 10 $\mu$m, which is an order of magnitude greater than the macroscopic estimation of Volovik, based on the same model. Because of inaccuracy of estimations and an approximate character of the model one can not consider this as a serious disagreement between the two estimations. The value 1 $\mu m$ better explains the NMR data \cite{dm2} which indicate that $\xi_{LIM}$ has to be much smaller than the dipole length $\xi_D\sim 10 \mu m$. The apparent discrepancy of the present estimation of $\xi_{LIM}$ with experiment can be ascribed to the neglect of the effect of  correlations of positions and orientations of building elements of aerogel in evaluation of $K(\mathbf{k})$ Eq. (19). Presence of such correlations in the silica aerogel was  demonstrated directly by X-ray scattering  \cite{parp1}. Account of the correlations
 in evaluation of other physical quantities renders an extra factor factor $(R/\xi_0)^2$, where $R$ is a correlation radius  \cite{fom1}. In the 98\% silica aerogel this factor can rise estimation for $K(\mathbf{k})$  $\approx 6\div 9$ times. That will bring closer two estimations of $\xi_{LIM}$.

\section{}
Global anisotropy can occur in a process of preparation of a sample or be a result of deformation \cite{hlp}. For a formal description of a global anisotropy the additional term of the form $\kappa_{jl}\Delta_j\Delta^*_l$ has to be added to the density of free energy Eq. (2). Here $\kappa_{jl}$ is a constant symmetric traceless tensor. As a consequence the equations of equilibrium acquire additional terms.  Eq. (4) does not change, but Eqns. (5) and (6) now look as:
$$
\mathbf{l}\cdot(\overrightarrow{D\mathbf{m}})=\mathbf{m}\cdot(\overrightarrow{\eta\mathbf{l}})+
\mathbf{m}\cdot(\overrightarrow{\kappa\mathbf{l}}),       \eqno(18)
$$
$$
\mathbf{l}\cdot(\overrightarrow{D\mathbf{n}})=\mathbf{n}\cdot(\overrightarrow{\eta\mathbf{l}})+
\mathbf{n}\cdot(\overrightarrow{\kappa\mathbf{l}}).       \eqno(19)
$$
In  zero-order approximation over  $\eta_{jl}(\mathbf{r})$   Eqns. (4),(18),(19) have spatially uniform solutions determined by the conditions $\mathbf{m}\cdot\overrightarrow{\kappa\mathbf{l}}=0$ and $\mathbf{n}\cdot\overrightarrow{\kappa\mathbf{l}}=0$.
These conditions show that in equilibrium  $\mathbf{m}_0,\mathbf{n}_0,\mathbf{l}_0$ have to be aligned with the principal directions   of $\kappa_{jl}$ which can be denoted as $\mathbf{u},\mathbf{v},\mathbf{w}$ with the corresponding principal values $\kappa_u,\kappa_v,\kappa_w$. As before, $\mathbf{m}_0,\mathbf{n}_0,\mathbf{l}_0$ are used as a basis of coordinate system $\hat{x},\hat{y},\hat{z}$.
A gain of orientational energy is proportional to $\sim\kappa_{jl}l_jl_l$. The energy is minimized when
$\mathbf{l}$ is either parallel or anti-parallel to the principal direction of $\kappa_{jl}$ corresponding to the maximum principal value. Both $\kappa_y-\kappa_z$ and $\kappa_x-\kappa_z$ are not positive.

 We consider in detail an axially symmetric aerogel when two of the three principal values coincide. Then all three can be expressed in terms of one parameter $\kappa$: $\kappa_u=\kappa_v=\kappa$ and $\kappa_w=-2\kappa$. There are two possibilities: 1)$\kappa<0$, then the
maximum principal value $\kappa_w$ is not degenerate. Such anisotropy is produced by a uniaxial compression of isotropic aerogel.
 In this case $\kappa_x-\kappa_z=\kappa_y-\kappa_z=3\kappa$ and a procedure, used at derivation of Eq. (13) renders
$$
\frac{\langle[\delta\mathbf{l}(\mathbf{r})]^2\rangle}{\mathbf{l}_0^2}=\int[1-\exp(i\mathbf{k}\cdot\mathbf{r})]
\frac{K(\mathbf{k})}{\xi_s^4(p^2+k_x^2+k_y^2+
\gamma^2k_z^2)^2}\frac{d^3k}{(2\pi)^3},                                 \eqno(20)
$$
where $p^2=3|\kappa|/\xi_s^2$. In case of $\mathbf{k}$-independent $K(\mathbf{k})$  $\rho$-dependence of $\langle[\delta\mathbf{l}(\mathbf{r})]^2\rangle$ can be written explicitly:
$$
\frac{\langle[\delta\mathbf{l}(\mathbf{r})]^2\rangle}{\mathbf{l}_0^2}=\frac{1}{\xi_{LIM}}\frac{1-e^{-p\rho}}{p}   \eqno(21).
$$
For small distances $p\rho\ll 1$ we are back to a linear growth of $\langle[\delta\mathbf{l}(\mathbf{r})]^2\rangle$ with a distance $\rho$ as described by the Eq. (16), but when  $p\rho\gg 1$ the ratio  $\frac{\langle[\delta\mathbf{l}(\mathbf{r})]^2\rangle}{\mathbf{l}_0^2}$ tends to a constant value
$D\equiv\frac{\xi_s}{\xi_{LIM}\sqrt{3|\kappa|}}$. If $D\ll 1$ (or $3|\kappa|\gg\left(\frac{\xi_s}{\xi_{LIM}}\right)^2$)
 the linear approximation applies for all $\rho$.  A quantity, which can be compared with the NMR data is $\langle l_z^2\rangle$. It enters an overall factor $q=\frac{3}{2}\left(\langle l_z^2\rangle-\frac{1}{3}\right)$ in the expression for the NMR frequency shift \cite{dm2}.
In terms of $\vec{\theta}$ the factor $q=1-\frac{3}{2}(\langle\theta_x^2\rangle+\langle\theta_y^2\rangle)$. Repeating calculations preceding Eq. (13) we arrive at
$$
\langle\theta_x^2\rangle+\langle\theta_y^2\rangle=\int\frac{K(\mathbf{k})}{\xi_s^4(p^2+k_x^2+k_y^2+
\gamma^2k_z^2)^2}\frac{d^3k}{(2\pi)^3}.                                 \eqno(22)
$$
At $\mathbf{k}$-independent $K(\mathbf{k})$    $\langle\theta_x^2\rangle+\langle\theta_y^2\rangle=D$.
If the anisotropy is induced by deformation of aerogel $\kappa\sim\epsilon$, where $\epsilon$ is a parameter, characterizing a uniaxial deformation. E.g. for a simple compression along z-axis $\epsilon=\delta h_z/h_z$. The dependence of $q$ on $\epsilon$ is then: $q=1-\sqrt{\frac{\epsilon_c}{\epsilon}}$. Characteristic deformation $\epsilon_c=\frac{3\epsilon}{4|\kappa|}\left(\frac{\xi_s}{\xi_{LIM}}\right)^2$ depends on a structure of aerogel. Within the ``random cylinders'' model for a simple compression $\epsilon_c=\frac{2\cdot7\cdot\zeta(3)}{\pi^2(1-P)}\frac{\varrho\xi_0}{\xi_{LIM}^2}$. For 98\% aerogel with $\varrho\approx$2 nm and $\xi_0\approx$20 nm we arrive at $\epsilon_c\approx 10^{-2}\div10^{-3}$ as in Ref. \cite{vol1}. The OF state is stabilized if $\epsilon\gg\epsilon_c$. At $\epsilon\approx\epsilon_c$  $\langle l_z^2\rangle\approx(1/3)$ and the OG state is realized.

In the case 2) ($\kappa>0$) each of the coinciding principal values is greater than the third value. Vector  $\mathbf{l}$ belongs to the $u,v$-plane, e.g. is parallel to $u$-axis, then $\kappa_{z}=\kappa_{y}\equiv\kappa$. The third principal value is   $\kappa_{x}=-2\kappa$.  With the same model assumption about the correlation function in this case:
$$
\frac{\langle[\delta\mathbf{l}(\mathbf{r})]^2\rangle}{\mathbf{l}_0^2}=\frac{1}{2\xi_{LIM}}\left(\rho+\frac{1-e^{-p\rho}}{p}\right)   \eqno(23).
$$
At $p\rho\gg 1$ the first term in the brackets dominates and a glass state is realized, but it is different from the state, described by Eqns. (7)-(9). A straightforward calculation shows that
$\langle l_x^2\rangle/l_0^2=\langle\theta_y^2\rangle=\frac{D}{2}$. If $D\ll 1$ $\mathbf{l}$ varies mainly in the $yz$-plane, or with respect to
$\mathbf{u},\mathbf{v},\mathbf{w}$ axes in the $u,v$-plane. This state is referred as a planar LIM state \cite{dm2}.
The above argument can be easily reformulated for a situation when all three principal values of $\kappa_{jl}$ are different.

The proposed linear theory relates statistical properties of random orbital textures in $^3$He-A to correlation functions of the random anisotropy created by the strands of aerogel. Here the procedure was illustrated with the aid of simple model of aerogel, which admits an analytical treatment.  Accuracy of the obtained numerical evaluations of characteristic parameters of textures depends on a model and it can be improved by choice of a more realistic model or by a direct use of experimental data for correlations in aerogel.
More restrictive is requirement imposed on distances $r\ll\xi_{LIM}$. It means that the theory describes local properties of random textures. Relation between local and global properties of glass-like states of $^3$He-A and analysis of their superfluid properties  deserve further investigation.

I thank V.V.Dmitriev for useful discussions and comments.
This work was supported in part by the Russian Foundation for Basic Research, project \# 14-02-00054-a

\end{document}